\documentclass[twocolumn,prc,nofootinbib,showpacs,epsfig]{revtex4}
\usepackage{graphicx}
\usepackage{epsfig}

\newcommand{\act}{$^{228}$Ac}
\newcommand{\tho}{$^{232}$Th}
\def\Journal#1#2#3#4{{#1} {\bf #2}, #3 (#4)}

\def\NPA{{\em Nucl. Phys.} A} 
\def\NP{\em Nucl. Phys.}

\def\PLB{{\em Phys. Lett.} B}

\def\RMP{{\em Rev. Mod.  Phys. }}

\def\PR{\em Phys. Rev.} 
\def\NAF{\em Z. Naturforschung A} 
\def\RPP{\em Rep. Prog. Phys.}
\def\PHM{\em Philos. Mag.} 
 
\def\PRC{{\em Phys. Rev.} C}

\def\ZP{\em Z. Phys.}

\def\ARI{\em Appl. Rad. Isot.}
\def\PAC{\em Pure Appl. Chem.}
\newcommand{\be}{\begin{equation}}
\newcommand{\ee}{\end{equation}}
\def\bea{\begin{eqnarray}} 
\def\eea{\end{eqnarray}}

\newcommand{\vf}{\mbox{$^{50}$V }}
\newcommand{\tf}{\mbox{$^{50}$Ti }} 
\newcommand{\cf}{\mbox{$^{50}$Cr }}

\begin{document}
\flushbottom
\title{Precision half-life measurement of the 4-fold forbidden electron capture of \vf}
\author{H. Dombrowski$^{a}$, S. Neumaier$^{a}$ and K. Zuber$^{b}$}

\affiliation{
$^{a}$Physikalisch-Technische Bundesanstalt (PTB), 38116 Braunschweig, Germany\\
$^{b}$Institut f\"ur Kern- und Teilchenphysik, Technische Universit\"at Dresden, 01069 Dresden, Germany
}
\begin{abstract}
A sensitive search of the 4-fold forbidden non-unique beta decay of \vf has been performed. A total exposure of 185.8 kg $\cdot$ d has been accumulated. A reliable half-life value with the highest precision so far of $(2.29 \pm 0.25) \cdot 10^{17}$~years of the electron capture decay of \vf into the first excited state of \tf could be obtained. A photon emission line following the 4-fold
forbidden beta decay into the first excited state of \cf could not be observed, resulting in a lower limit on the half-life of the beta decay branch of $1.7 \cdot 10^{18}$~years. This is barely in agreement with a claimed observation of this decay branch.
\end{abstract}
{\small P
ACS: 13.15,13.20Eb,14.60.Pq,14.60.St}


\maketitle
\section{Introduction}
\begin{figure}
  \centering
  \includegraphics[width=80mm]{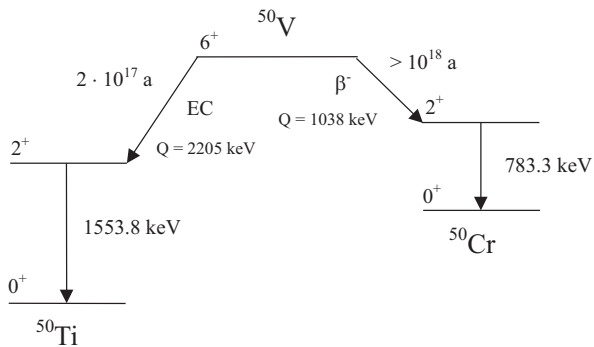}
\caption{Decay scheme of $^{50}$V. Two excited states can be populated, one via electron capture to \tf under emission of a 1553.8 keV gamma and the beta decay into the first excited state of \cf resulting in a 783.3 keV gamma.}
  \label{pic:levelvf}
\end{figure}

In the history of particle and nuclear physics the study of weak interactions and especially of beta decay has played a vital role. Those studies helped, among others, to establish the V-A structure of weak
interactions. Nowadays, this interest in exploring beta decays and related issues 
is rather reduced but still there are interesting topics
to investigate like the endpoint measurements of tritium and $^{187}$Re electron spectra to determine
the neutrino mass \cite{ott08} or to search for S,T,V contributions to the weak interaction \cite{bec06}.
In addition to these ``beyond the standard-model'' searches also interesting nuclear physics questions are 
still open which can be addressed by studying highly forbidden beta decays. 
The majority of beta emitters is characterized
as allowed or single forbidden, however there are a few isotopes in nature 
which are even at least 4-fold forbidden \cite{sin98}. 
Their decays are extremely rare with half-lives well beyond
$10^{13}$ years. Compared to those decays, which are at least 5-fold forbidden transitions (like $^{48}$Ca and $^{96}$Zr), even a double beta decay is more likely to happen.\\ \indent
In this paper the focus is on the 4-fold forbidden ${\beta}$-decay of $^{50}$V. 
There are only three nuclei in nature which permit a feasible study of 4-fold 
forbidden beta decay, namely $^{50}$V, $^{113}$Cd and $^{115}$In, all of them are non-unique ($\Delta I ^{\Delta \pi} = 4^+$). 
Half-lives of these transitions are long ($\geq10^{14}$ years) and would produce very low count rates in typical experiments. 
Such measurements can only be performed in well shielded facilities with a low radioactive background.  
Recent measurements of half-lives of $^{113}$Cd are reported in \cite{goe05,bel07,daw09} and of $^{115}$In
in \cite{pfe79,cat05}, the latter including also a transition into the first excited state. These activities triggered 
for the first time theoretical attempts to calculate the energy spectra  of such beta decays \cite{mus06,mus07}.\\ \indent
The isotope \vf is quite unique in the sense that in contrast to $^{113}$Cd and $^{115}$In
the ground state transition is even higher-forbidden, leaving only 4-fold forbidden non-unique decay modes 
into the first excited states of \cf and $^{50}$Ti, both characterized as 6$^+ \rightarrow 2^+$ transitions. 
The ground state transitions to both isotopes are even 6-fold forbidden non-unique decays. The decay
scheme is shown in Fig.~\ref{pic:levelvf}. 
The Q-value for the beta decay into \cf is (1037.9~$\pm$~0.3)~keV  and for electron capture (EC) into \tf
it is (2205.1~$\pm$~1)~keV, respectively \cite{wap03}. There is only one excited state in each daughter nucleus 
which can be populated. The corresponding gamma lines to search for are 1553.77~keV for EC into the first excited state of \tf and 783.29~keV for the beta decay into the first excited state of $^{50}$Cr, respectively. The photon emission probability of both E2 transitions is 1 (with a negligible uncertainty). \\ \indent
Various attempts have been made to observe the decay of \vf into \tf \cite{hei55,glo57,bau58,mcn61,wat62,son66,pap77,alb84,sim85,sim89}. However, the deduced half-life 
changed by several orders of magnitude over the decades, while uncertainties claimed were typically well beyond 20 \%. 
For the beta decay into the excited state of \cf only lower
limits were published except one vague indication of an observation \cite{sim89}.
The measurements within the last 45 years are compiled in Tab.~\ref{tab:exppara}.
All these measurements were performed more than at least two decades ago. 

\begin{table}[htdp]
\begin{center}
\begin{tabular}{|c|c|c|c|c|}
\hline
Mass   & Time       & $T_{1/2}^{EC}$       & $T_{1/2}^{\beta^-}$              & Ref.\\ 
g      & d          & $10^{17}$a           & $10^{17}$a                       &     \\
\hline \hline
4000   &   48.88    &  $>8.8$              & $>7.0$                           & \cite{pap77}\\ 
4250   &   135.5    &  $1.5^{+0.3}_{-0.7}$ & $>4.3$                           & \cite{alb84}\\ 
100.6  &   8.054    &  $1.2^{+0.8}_{-0.4}$ & $>1.2$                           & \cite{sim85}\\ 
337.5  &   46.21    &  $2.05 \pm 0.49$     & $8.2^{+13.1}_{-3.1}$             & \cite{sim89}\\
255.8  &   97.8     &  $2.29 \pm 0.25$     & $>15$                            & this work  \\ 
\hline
\end{tabular}
\caption{\label{tab:exppara} Measurements of \vf decays in the last 45 years.}
\end{center}
\end{table}
The aim of this paper is to perform a high statistics search and provide a precision measurement of the \vf decay taking advantage of a sophisticated detector system and also of developments in low background techniques.

\section{Experimental procedure}
The measurement was performed using an ultra-low background Ge-detector (ULB detector) at the underground laboratory for dosimetry and spectrometry of the PTB (UDO) close to Braunschweig (Germany). UDO is located inside the Asse salt mine providing shielding with respect to the secondary cosmic radiation of about 1200~m of water equivalent. The p-type extended range ULB detector has an efficiency of 89~\% and an energy resolution of 2.1~keV at 1.33~MeV. The mass of the Ge single crystal is 1.9~kg. It is surrounded by a shielding of low activity made of inner 10 cm of electrolytic copper and two outer layers of lead with a $^{210}$Pb content of 2~Bq/kg and 6~Bq/kg, respectively. For a detailed description of the apparatus see  \cite{neu09}. \\ \indent
The detector was calibrated by recording the spectra of cans filled with solutions of well known activity provided by PTB. The activity of single nuclides within these solutions is known within an uncertainty between  1~\% and 2~\%. The solutions covered an energy range of strong emission lines from 21.0~keV to 1836~keV, using 11 radionuclides producing 18 major emission lines. To obtain
the efficiency the sample-detector geometry was modeled by applying the Monte Carlo code GESPECOR (some geometrical parameters of the detector known from the technical drawing were altered slightly to achieve a better agreement between measurement and Monte Carlo simulation). Summing corrections were obtained by using the same code. The code served to calculate the efficiency of the detector concerning the \vf sample as well. An efficiency transfer was calculated from a 100 ml can containing a known radionuclide concentration to the same 100 ml can filled with vanadium powder (with a natural isotopic composition). This procedure led to an uncertainty of the efficiency of less than 3~\% in the covered energy region. More information on this detector and its performance can be found in \cite{neu09}. \\ \indent
The uncertainties published in this article were calculated by adding all known uncertainty contributions  quadratically to obtain the value of the total uncertainty. This procedure is in agreement with the Guide to the expression of uncertainty in measurement, GUM \cite{gum08}. The uncertainties in this article are expressed as standard uncertainties (coverage factor $k = 1$).

\section{Experimental details}
The background of the applied detector in the region of the 1553.8~keV line, integrated from 1553~keV to 1555~keV, is lower than 0.025 counts per day. The pure background in the region of the 783.3 keV line is also rather low: Integrated from 782 keV to 784 keV, it is about 0.25 counts per day. But thorium and uranium impurities of the sample lead to a much higher background in the region of the 783.3 keV line (see below), while there are no natural lines near the 1553.8 keV line. 
As the measurement is completely dominated by contaminations within the sample, the intrinsic background of the 
set-up can be neglected.\\ \indent
A sample of vanadium powder with a total mass of 255.82~g, corresponding to a volume of 100 ml, was filled in a cylindrical film can (made of plastic) and placed on top of the above mentioned detector.
The total  measuring time was 97.8~days resulting in a total detector mass times measuring time product of 185.8 kg d. The measurement confirmed that the sample was not highly purified because various emission lines from the natural decay chains of uranium and thorium are visible, caused by activities in the mBq range. \\ \indent
Vanadium is a base metal, hence the surface will oxidize when it has contact with air. In addition, it can absorb some water. The can in which the vanadium powder was kept was not completely hermetically sealed, so that it had contact with the surrounding air. The oxygen content of the sample of ($6.13 \pm 0.12$)~\% by mass was measured by the German Federal Institute for Materials Research and Testing (BAM) with a high precision. The water content of ($1.445 \pm 0.010)$~\% by mass was determined by PTB by drying the sample at low temperatures ($< 100^\circ$ C). The mass of the oxygen and water was subtracted from the mass of the total sample before calculating the activity concentration and half-lives of vanadium.

\section{Results}

\begin{figure} 
  \centering
        \includegraphics[width=\linewidth]{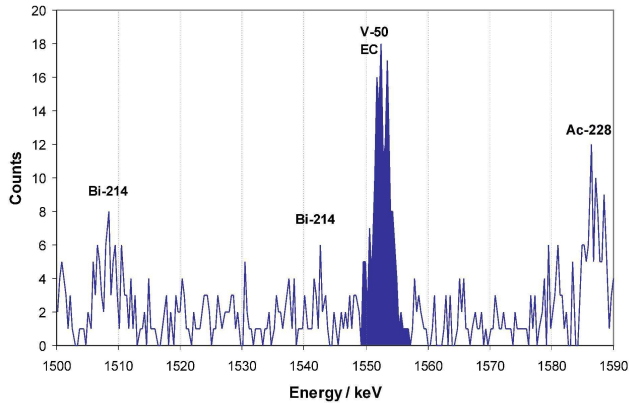}
  \caption{Energy spectrum in the range from 1500~keV to 1590~keV. The \vf EC decay line is marked in grey.}
  \label{pic:highEline}
\end{figure}

\subsection{The 1553.8 keV line from electron capture}
The energy range around the expected line from EC is shown in Fig.~\ref{pic:highEline}. The region shows two prominent lines, one at the expected energy and one at 1588 keV. The latter is emitted by \act, a progeny of \tho \ (nuclear data information taken from \cite{nudat}). A fit to the first line results in 123 net counts. By taking into account the detection efficiency of 2.17~\% and the natural abundance of \vf of 0.25~\% \cite{lae03} this can be converted into a partial half-life of $(2.29 \pm 0.25) \cdot 10^{17}$~years. The fact that the 1554.1 keV line from $^{234m}$Pa (from the $^{238}$U series) could interfere with the line of interest has to be taken into account. This $^{234m}$Pa line is in equilibrium with a gamma line at 1001.1 keV, which should include 139 times more counts because of a higher emission probability and a higher efficiency of the detector at that energy. As the line at 1001.1 keV is not visible at all in the measured spectrum, the $^{234m}$Pa line contribution at 1554.1 keV is neglected. \\ \indent
All known sources of uncertainty bigger than 0.001\%, which have an influence on the calculated life-time, are listed in Table \ref{tab:uncert}. The total uncertainty of the half-life of the EC decay is clearly dominated by the uncertainty of the determination of the number of detected counts (i.e. fitting of the 1553.8 keV line and its background).

\begin{table}[htd]
\begin{center}
\begin{tabular}{|l|c|}
\hline
Contribution                                     & Uncertainty \\
																								 & in \% \\
\hline
\hline
$^{50}$V abundance in sample (total)             & \hspace{-5mm}2.5 \\
including:                                       & \\
\hspace{3mm} Natural $^{50}$V isotopic abundance & \hspace{5mm}1.6 \\
\hspace{3mm} V concentration due to manufacturer & \hspace{5mm}0.1 \\
\hspace{3mm} Water content of sample             & \hspace{5mm}0.7 \\ 
\hspace{3mm} Oxygen content of sample            & \hspace{5mm}2.0 \\
Weight of sample                                 & \hspace{-5mm}0.04 \\
Activity determination                           & \hspace{-5mm}10 \\
including:                                       & \\
\hspace{3mm} Number of detected counts           & \hspace{5mm}10 \\
\hspace{3mm} Detector efficiency                 & \hspace{5mm}3  \\
&\\
Total uncertainty of half-life                   & \hspace{-5mm}11 \\
\hline    
\end{tabular}
\caption{\label{tab:uncert} Contributions to standard uncertainty assigned to the half-life of the EC decay of $^{50}$V. Contributions smaller than 0.001~\% are not listed (especially the atomic weight of vanadium as well as the measuring time are known with a much higher precison).}
\end{center}
\end{table}

\begin{figure}
\centering
\includegraphics[width=\linewidth]{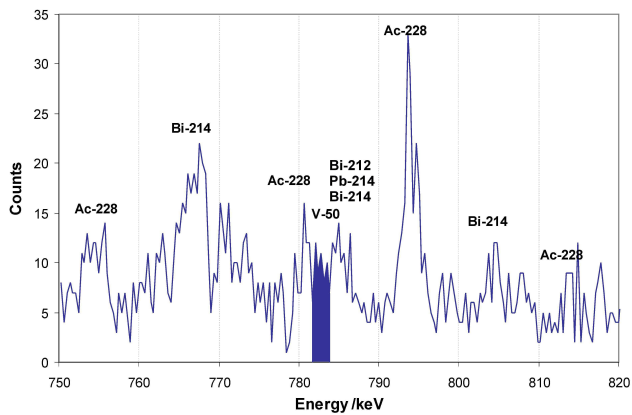}
\caption{Energy spectrum from 750~keV to 820~keV. The region, where the beta decay line of \vf is expected, is marked in grey.}
\label{pic:lowEline}
\end{figure}

\subsection{The 783.3 keV line from beta decay}
Many more lines are visible in the range of the potential $\gamma$-line from the beta decay into the first 
excited state of \cf (Fig.~\ref{pic:lowEline}). At around 783~keV there are six known lines from the U and Th
chains: 782.1~keV ($^{228}$Ac), 783.4~keV ($^{234m}$Pa), 785.4~keV ($^{212}$Bi), 786.0~keV ($^{214}$Pb), 786.1~keV ($^{214}$Bi) and 786.3~keV ($^{234m}$Pa) \cite{nudat}. The contributions of the $^{234m}$Pa lines can be neglected for similar reasons that were explained in the section before. But in direct proximity to the energy of interest, 783.3~keV, the contributions of the remaining lines form peak-like structures. A line at 783.3~keV is not visible. On the contrary, a dip is found at that energy. An upper detection limit of 27 events (using a significance level of $\alpha=5~\%$) for a potential line can be calculated according to the ISO 11929 standard\cite{din11}, a well defined standard procedure for sample measurements. With a detection efficiency of 3.35~\% this converts into a lower limit on the partial half-life of the beta decay $1.5 \cdot 10^{18}$~years. This is larger than the claimed observation in \cite{sim89}. If the partial half-life published in \cite{sim89} had been correct, a total of 39 events should have been observed in our measurement, which is excluded by two standard deviations, if the large upper error of 160~\% stated in \cite{sim89} is not taken into account. 
The lower limit of the half-life of the beta decay published in this article shows that the value of $8.2 \cdot 10^{17}$ years entering current nuclear tables (like \cite{nudat}) is too small, as the new value of the lower limit is nearly twice as high. 
\\ \indent
The uncertainty budget of our evaluation of the 783.3~keV line is very similar to that of our evaluation of the 1553.8~keV line, but here an uncertainty resulting from the determination of the number of detected counts is not applicable. As the evaluation of the region of the 783.3~keV line only results in a lower limit, this value is not combined with an uncertainty.

\section{Summary and Conclusions}
More than two decades since the last attempt, the 4-fold forbidden decays of \vf were investigated ahain with an ultra-low background Ge-detector system located deep underground. For the first time ever oxygen and water content of the vanadium
sample were determined to get a more accurate value for the real vanadium mass. The electron capture decay branch, which populates the first excited state of $^{50}$Ti, is clearly visible with a half-life of
\be
T_{1/2} (^{50}\rm{V}\rightarrow \rm{^{50}Ti} + 1553.8\,keV) = (2.29 \pm 0.25) \cdot 10^{17}\,\rm{a}
\ee
This is roughly a factor of two more precise compared to the best claim\cite{sim89}. 
The beta decay branch, which leads to the first excited state of $^{50}$Cr,
could not be observed and a lower half-life limit of 
\be
T_{1/2} (^{50}\rm{V}\rightarrow \rm{^{50}Cr}  + 783.3\,keV) > 1.5  \cdot 10^{18}\,\rm{a}\
\ee
is concluded (with a significance level of 5 \%). This is barely in agreement with the only positive claim made of this decay in the past. From both results of this article a combined total half-life of \vf with a lower limit of $2.0\cdot 10^{17}$ years can be derived. That means that 7.1~\% of the decays happen by beta decay at maximum. As a consequence, the branching ratio of the two decay modes of \vf is not correctly implemented in \cite{nudat}.

The current search was limited by an intrinsic contamination of the vanadium sample and of the detector. The actual background spectrum of the detector itself is about a factor of 4 lower concerning the $^{238}$U decay chains and a factor of 2 better concerning the $^{232}$Th decay chain. Thus, after purifying the vanadium, a follow-up measurement allows a more sensitive search for the beta decay mode of $^{50}$V.

\section{Acknowledgements}
The authors would like to express their gratitude to the following members of the PTB radioactivity department for their support of this work: Dirk Arnold, Marion Ehlers, Karsten Kossert, Michael Schmiedel and Herbert Wershofen. In addition, the authors are very grateful for the support of Heinrich Kipphardt and Nicole Langhammer, both members of BAM.


\begin{thebibliography}{9}
\bibitem{ott08} E.~W.~Otten, C.~Weinheimer, \Journal{\RPP}{71}{086201}{2008}
\bibitem{bec06} N.~Severijns, M.~Beck, O.~Naviliat-Cuncic,  \Journal{\RMP}{78}{991}{2006}
\bibitem{sin98} B. Singh et al., {\it Nucl. Data Sheets} 84, 487 (1998)
\bibitem{goe05} C.~G\"{o}ssling et al., \Journal{\PRC}{72}{064328}{2005}
\bibitem{bel07} P.~Belli et al., \Journal{\PRC}{76}{064603}{2007}
\bibitem{daw09} J.~Dawson et al., \Journal{\NPA}{818}{264}{2009}
\bibitem{pfe79} L~ Pfeiffer et al., \Journal{\PRC}{19}{1035}{1979}
\bibitem{cat05} C.~M.~Cattadori et al., \Journal{\NPA}{748}{333}{2005}
\bibitem{mus06} M.~T.~Mustonen, M.~ Aunola, J. Suhonen, \Journal{\PRC}{73}{054301}{2006}, err.  \Journal{\PRC}{76}{019901}{2007}
\bibitem{mus07} M.~T.~Mustonen, J. Suhonen, \Journal{\PLB}{657}{38}{2007}
\bibitem{wap03} G. Audi, A. H. Wapstra, C. Thibault, \Journal{\NPA}{729}{337}{2003}
\bibitem{hei55} J.~Heintze, \Journal{\NAF}{10}{77}{1955}
\bibitem{glo57} R.~N.~Glover, D.~E.~Watt, \Journal{\PHM}{2}{697}{1957}
\bibitem{bau58} E.~R.~Bauminger, S.~G.~Cohen, \Journal{\PR}{110}{953}{1958}
\bibitem{mcn61} A.~McNair, \Journal{\PHM}{6}{559}{1961}
\bibitem{wat62} D.~E.~Watt, R.~L.~G.~Keith, \Journal{\NP}{29}{648}{1962}
\bibitem{son66} C.~Sonntag, K.~O.~M\"unnich, \Journal{\ZP}{197}{300}{1966}
\bibitem{pap77}  A.~Pape, S.~M.~Refaei, J.~C.~Sens, \Journal{\PRC}{15}{1937}{1977}
\bibitem{alb84} D.~E.~Alburger, E.~K.~Warburton,  J.~B.~Cumming, \Journal{\PRC}{29}{2294}{1984}
\bibitem{sim85}  J.~J.~Simpson, P.~Jagam, A.~A.~Pilt, \Journal{\PRC}{31}{575}{1985}
\bibitem{sim89} J.~J.~Simpson, P.~Moorehouse, P.~Jagam, \Journal{\PRC}{39}{2367}{1989}
\bibitem{neu09} S. Neumaier et al., \Journal{\ARI}{67}{726}{2009}
\bibitem{gum08} BIPM: Evaluation of measurement data - Guide to the expression of uncertainty in measurement, \Journal{JCGM}{100:2008} {www.bipm.org} {2008}
\bibitem{lae03} J.~R.~de Laeter et al., \Journal{\PAC}{75}{683}{2003}
\bibitem{nudat} C. L. Dunford and R. R. Kinsey, NuDat system for access to nuclear data. IAEA-NDS-205 (BNL-NCS-65687), July 1998, information extracted from the NuDat data base, version of 17 March 2004, using the PC version of the program NuDat (Vienna, Austria: IAEA) (2004)
\bibitem{din11} Deutsches Institut f\"ur Normung E.V. (German National Institute for Standards). Determination of the characteristic limits (decision threshold, detection limit and limits of the confidence interval) for measurements of ionizing radiation - Fundamentals and application (DIN ISO 11929:2011-01) (Berlin: DIN) (1993)




\end{thebibliography}
\end{document}